\documentclass{aa}
\usepackage{graphics}

\def\4261{\object{NGC\,4261}}
\begin{document}

   \thesaurus{1 
              (11.01.2;  
               11.09.1 NGC\,4261;  
               13.19.1)}  
   \title{A thin H\,{\sc i} circumnuclear disk in \4261}

   \author{H.J.~van Langevelde
          \inst{1}
          \and
          Y.M.~Pihlstr\"om\inst{2}
          \and
          J.E.~Conway\inst{2}
          \and
          W.~Jaffe\inst{3}
          \and
          R.T.~Schilizzi\inst{1,3} }

   \offprints{YMP ({\tt ylva@oso.chalmers.se})}

   \institute{Joint Institute for VLBI in Europe, Postbus 2, 7900 AA, Dwingeloo, The Netherlands
         \and
             Onsala Space Observatory, S-439 92 Onsala, Sweden
         \and
             Sterrewacht Leiden, Postbus 9513, 2300 RA, The Netherlands
             }

   \date{Received ; accepted }

   \maketitle

   \begin{abstract} We report on high sensitivity, spectral line VLBI
     observations of the H\,{\sc i} absorption feature in the radio
     galaxy \4261.  Although absorption is only detectable on the most
     sensitive baseline, it can be unambiguously associated with the
     counterjet and is interpreted to originate in a thin atomic
     circumnuclear disk. This structure is probably a continuation of
     the dusty accretion disk inferred from HST imaging, which could
     be feeding the massive black hole. H\,{\sc i} column densities in
     front of the counterjet of the order of $10^{21}(T_{\rm
     sp}/100\,{\rm K})$ cm$^{-2}$ are derived, consistent with X-ray
     data and VLBI scale free-free absorption.  The data presented
     here are the result of the first scientific project processed on
     the new EVN MkIV data processor.
     
\end{abstract}
%

\section{Introduction}

Perhaps the most striking feature of the FRI radio galaxy \4261
(3C\,270) is its approximately 240 pc radius circumnuclear dust disk
revealed by HST observations (Jaffe et al.\ \cite{jaffe93}).  Further
optical observations of the kinematics of emission lines in the inner
regions of this disk by Ferrarese et al.\ (\cite{ferrarese96}) have
since provided evidence for a central black hole of mass of
$4.9\times10^8$M$_{\odot}$. In the radio band, H\,{\sc i} absorption
has been detected toward the core of \4261 using the VLA (Jaffe \&
McNamara \cite{jaffe94}). It was argued by these authors that this
absorption is due to atomic hydrogen in the inner part of the HST
disk. Such a disk interpretation is consistent with high resolution
VLBI or MERLIN H\,{\sc i} absorption observations in a number of other
AGN. For example VLBA H\,{\sc i} observations of another FRI galaxy,
Hydra A, are consistent with a 20 pc flattened disk structure (Taylor
\cite{taylor96}). In this Letter we report on VLBI observations of
\4261 performed using the high-sensitivity antennas of the European
VLBI Network (EVN). The aims were to confirm that the H\,{\sc i} is
indeed associated with the HST dust disk and to better constrain the
disk geometry and physical properties.

Detailed studies of the dynamics and chemistry of circumnuclear disks
such as that found in \4261 are important for several reasons.
Such disks almost certainly provide the fuel which powers AGN, but the
accretion process is poorly understood. In addition such flattened
circumnuclear structures are required by orientation-based unified
schemes. While the inner edges of these occulting structures must be on
BLR scales (0.1 -- 1 pc) their outer radii are poorly defined and may
extend to hundreds of parsecs. Evidence for circumnuclear gas on a
variety of scales in different physical states has been accumulating.
Examples include HST imaging of parsec scale ionised gas in M87 (Ford
et al.\ \cite{ford94}), 100 -- 1000 pc scale molecular CO in Centaurus A
(Rydbeck et al.\ \cite{rydbeck93}) and HCN in the Seyfert 2
NGC\,1068 (Tacconi et al.\ \cite{tacconi94}). Amongst all the objects
observed \4261 is unique in showing optical dust, H\,{\sc i}
absorption and VLBI scale free-free absorption (Jones \& Wehrle
\cite{jones97}), allowing us to study the disk on a variety of scales.

The HST optical imaging of \4261 provides strong constraints on the
disk geometry at 100 pc scales. The ratio of the apparent major and
minor axes (Ferrarese et al.\ \cite{ferrarese96}, Jaffe et al.\
\cite{jaffe96}) implies, if the disk is circular, that its normal is
inclined $64^{\circ}$ to the line of sight. Modeling of the dust
obscuration shows that it is the East side of the disk which is
closest to us. Such modeling also shows that the dust disk is thin,
with a thickness $<40$ pc at its outer edge (Jaffe et al.\
\cite{jaffe96}). The normal to the disk is found in projection to be
roughly oriented along the radio axis, making an angle of $14^{\circ}$
to the kiloparsec scale radio jets. At both 1.6 and 8.4 GHz VLBI
observations show a two-sided jet in the same position angle as the
kiloparsec jets (Jones \& Wehrle \cite{jones97}). The Eastern jet,
which is slightly weaker, is assumed to be the counterjet, given that
the Eastern edge of the HST dust disk is tilted toward us and the
radio jets are roughly aligned along the disk axis. Consistent with
this orientation, Jones \& Wehrle (\cite{jones97}) argue that a narrow
gap in the 8.4 GHz radio emission toward the Eastern jet is from
free-free absorption via occultation by an inner ionised accretion
disk of radius 0.2 pc. In the remainder of the paper we discuss our
H\,{\sc i} VLBI observations and the additional constraints on disk
geometry and physical properties they provide. Throughout this paper
we assume a distance of $\sim30$ Mpc (Nolthenius \cite{nolthenius93}),
so 1 mas corresponds to 0.14 pc.

\section{Observations and Processing}\label{observations}
Observations of \4261 were made at 21 cm with the European VLBI
Network (EVN) on February 22 1999 and lasted 10 hours. Participating
antennas were Effelsberg, Jodrell Bank (Lovell telescope), Medicina,
Noto, Onsala, Torun, and Westerbork (phased array). Unfortunately one
of the three large collecting areas, the 100m at Effelsberg, was
unable to observe due to heavy snowfall. The observing mode consisted
of 4 frequency bands, all observed with dual circular polarisation, 4
MHz wide and 2-bit sampled. The second frequency band was centred on
the H\,{\sc i} line of \4261 at a velocity of 2237 km/s (heliocentric,
optical definition) corresponding to a frequency of 1410 MHz.

The data were processed on the EVN MkIV data processor at the Joint
Institute for VLBI in Europe (JIVE) and constitute the first
scientific experiment to be carried out with this new facility. The
data were correlated between July 26 and August 6 1999 in two passes;
the first resulting in 128 spectral channels on both polarisations of
the line data, the second pass yielding sensitive continuum data, by
processing all basebands with modest spectral resolution. For the
spectral line dataset the resulting spectral resolution from uniform
weighting was $\approx 7.4$ km/s. Data quality was good, except for
some bad tape passes from Torun and a large fraction of the data for
Medicina which was corrupted by interference. Amplitude calibration
was carried out in the standard way using the $T_{\rm sys}$ values and
gain curves from the stations. The continuum was imaged using standard
self-calibration and CLEAN deconvolution methods. From variations in
amplitude gain factor in the final step of amplitude self-calibration
we estimate the uncertainties in the overall flux density scales of
our images to be of order 15\%.

\begin{figure}
\resizebox{\hsize}{!}{\includegraphics{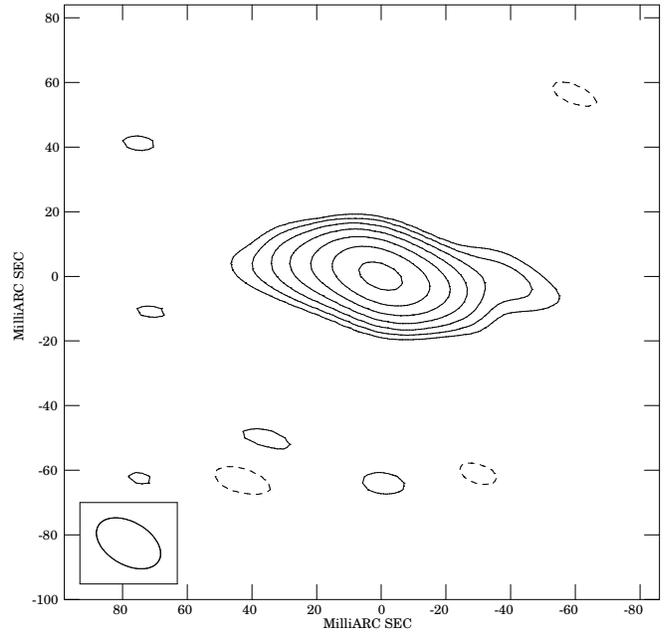}}
   \caption{Continuum image of \4261 at 21cm obtained with the
     EVN. The beam size is $21\times15$ mas. The contours are plotted
     at levels 1, 2, 4, 8, 16, 32 and 64 mJy/beam, with a map peak intensity
     of 79 mJy/beam. The Eastern counterjet side is only slightly weaker at
     this frequency.}
   \label{continuum}
\end{figure}

\section{Results}\label{results}
The continuum image obtained of \4261 is shown in Fig.\
\ref{continuum}. The noise level in this image is 0.35 mJy/beam. We
find that the Western jet is somewhat brighter and more extended than
the Eastern one, consistent with the VLBA maps of Jones \& Wehrle
(\cite{jones97}). From these VLBA maps it is clear that the flat
spectrum core lies close to the peak of Fig. \ref{continuum}. We were
able to fit a three component Gaussian model to the continuum
visibility data, consisting of one compact core component and two jet
components 18 and 14 mas (2.5 and 2 pc) to the East and West
respectively.

\begin{figure*}
\resizebox{12cm}{!}{\includegraphics{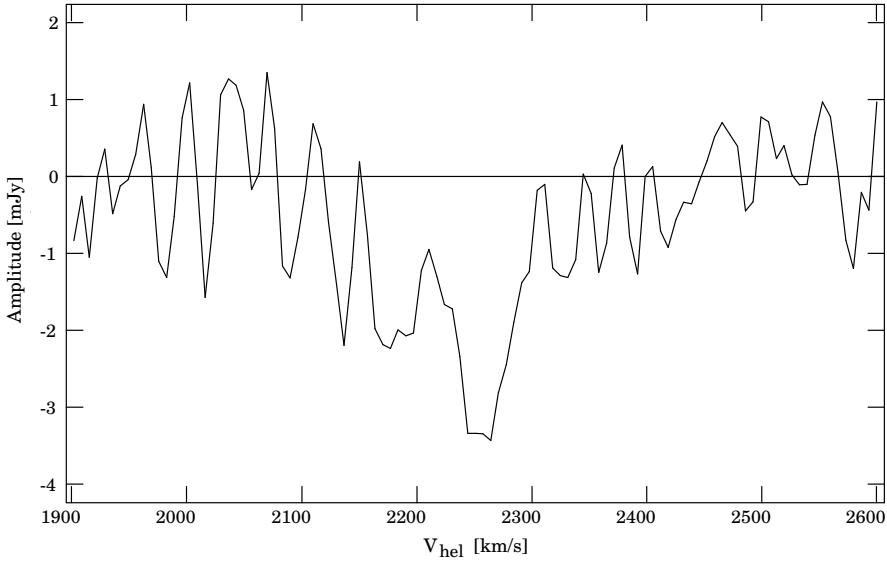}}
\hfill
\parbox[b]{55mm}{ \caption{Absorbed flux density for the Jodrell Bank
-- Westerbork baseline integrated 18 mas to the East of the map
centre, i.e.\, at the counterjet side. The continuum has been
subtracted.  One spectral channel corresponds to 27 km/s.}
\label{line}}
\end{figure*}

In order to detect the weak H\,{\sc i} absorption the spectral line
data was self-calibrated with the continuum image, and then the
continuum was subtracted using the AIPS task UVLIN. The spectral
absorption was unambiguously detected on the Jodrell Bank --
Westerbork baseline (Fig. \ref{line}). Other baselines did not have
sufficient sensitivity to give any detections. From the phase
information (not shown) it is clear that the absorption is not centred
on the reference position of the self-cal process; but is offset from
the core. The sign of the phase on the Jodrell Bank -- Westerbork
baseline suggests that the absorption is preferentially on the Eastern
(counterjet) side. Fig. \ref{line} shows the absorbed flux density
integrated at the supposed position of the main counterjet component
located 18 mas to the East of the core (see below).

From the VLA spectrum presented in the Jaffe \& McNamara
(\cite{jaffe94}) paper, we estimate a total integrated absorbed flux
density of $604\pm168$ mJy km/s, compared to the corresponding number
for our VLBI spectrum; $409\pm55$ mJy km/s. The amount of VLBI scale
absorption is therefore consistent with the VLA observations. Although
we cannot exclude the possibility of additional H\,{\sc i} absorbing
gas on scales larger than sampled by the VLBI observations, we feel
confident that we detect the bulk of the absorbing gas.

\begin{figure}
\resizebox{\hsize}{!}{\includegraphics{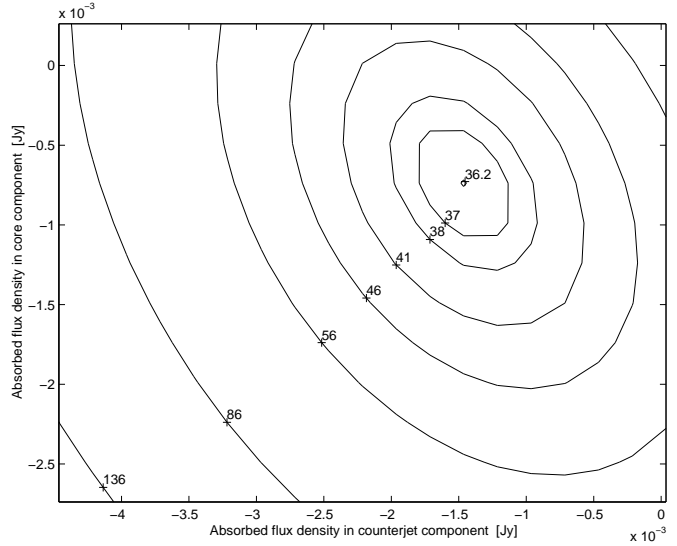}} \caption{The
     $\chi^2$-landscape achieved when varying the amount of absorbed
     flux over the counterjet and the core. The best fitting point has
     $\chi^2=36.2$. Number of degrees of freedom=35.}  
\label{chi2}
\end{figure}

In making quantitative estimates of the opacity toward different source
components we applied a model-fitting technique based on the three
component model used to fit the continuum data. We first averaged the
Jodrell -- Westerbork spectral absorption data in frequency over the
line width and then fitted the resulting phase and amplitude versus
time using a three component Gaussian model based on the continuum
model. Each Gaussian had the same fixed shape and position as that
fitted to the continuum data, only the amplitude of each component was
allowed to vary. The minimised $\chi^2$ is achieved when most of the
absorption is on the counterjet, a possible small absorption at the
core and no absorption against the jet component. Fixing the jet
absorption at zero we obtained the $\chi^2$-landscape shown in Fig.\
\ref{chi2} for different combinations of counterjet and core
absorption. From this we estimate the absorbed counterjet and core
flux densities averaged over the line to be $1.5\pm0.3$ mJy and
$0.7\pm0.4$ mJy respectively.

Dividing by the continuum flux densities of each component from the
absorbed fluxes we can estimate line-averaged opacities of
$0.11\pm0.02$ and $0.01\pm0.01$ against the counterjet and core
respectively. It therefore appears that virtually all of the absorbing
gas is against the counterjet. Integrating the opacity over the line
we estimate a total H\,{\sc i} column density towards the counterjet
of $N_{\rm HI} = 2.5\times 10^{19} \, T_{\rm sp}\ {\rm cm}^{-2}$ and
$N_{\rm HI} <2.2\times 10^{18} \, T_{\rm sp}\ {\rm cm}^{-2}$ towards
the core.

\section{Discussion}\label{discus}
Having most of the H\,{\sc i} column located in front of the
counterjet at a projected distance of $\sim2.5$ pc is consistent with
a number of previous results on \4261. First, at distances closer to
the nucleus we do not expect H\,{\sc i} absorption, since we know the
circumnuclear material is mostly ionised. This is shown by the
free-free absorption at a projected radius of 0.2 pc inferred by Jones
\& Wehrle (\cite{jones97}). Secondly, \4261 harbours an X-ray source
($\sim10^{41}$ erg/s in the 0.2-1.9 keV range, Worrall \& Birkinshaw
\cite{worrall94}), which puts a limit on the total column density on
the line of sight to the nucleus. The model presented by Worrall \&
Birkinshaw (\cite{worrall94}) yields an upper limit on the total
column density of $4\times10^{20} {\rm cm}^{-2}$. Given that the
X-rays preferentially originate in the nucleus, this fits in
comfortably with the constraints from our VLBI H\,{\sc i} absorption.

The model fitting from which we derive the optical depth does not
allow the positions of the components to vary (see Sect.\
\ref{results}), nor is there any sensitivity for H\,{\sc i} beyond the
end of the continuum counter-jet. We are therefore forced to make a
simplifying assumption, namely that 18 mas is the mean radius of the
H\,{\sc i} absorbing structure.  This is supported by the fact that
most of the VLA absorption is recovered by the VLBI observations
(Sect.\, \ref{results}). Given the HST dust disk inclination, this
implies a distance 5.7 pc away from the nucleus. The FWHM of the line
is comparable with other H\,{\sc i} absorption observations of
circumnuclear gas (e.g. in Cyg A, Conway \& Blanco \cite{conway95}).
Therefore, in the next step we assume that the atomic gas is part of
such a circumnuclear rotating structure and not due to individual
clouds randomly distributed in front of the continuum source. Such a
model of the H\,{\sc i} disk is supported by the nuclear parameters
derived by Ferrarese et al.\ (\cite{ferrarese96}) from HST data on
optical transitions. They found a central mass of $4.9\times10^8 {\rm
M}_{\odot}$, which implies a rotational velocity of 610 km/s at the
location of the H\,{\sc i}. Under the standard assumption that the
linewidth $\Delta V$ provides an estimate of the isotropic turbulent
velocity, we use the thin disk relation $h\sim r (\Delta V / V_{\rm
circ})$ to estimate the disk thickness $h$. We estimate the velocity
dispersion $\Delta V$ at radius $r$ to be $\sim130\, {\rm km/s}$ which
gives $h=1.3$ pc. So the H\,{\sc i} is likely to reside in a thin
circumnuclear disk with an opening angle of $\sim13^{\circ}$. The
average density can, assuming a volume filling factor $f$ of unity, be
estimated to be $n_{\rm HI} = 6\times10^2 {\rm cm}^{-3}$ for a spin
temperature of 100 K. It follows that a more clumpy distribution
($f<1$) will increase the estimated density ($\propto f^{-1/3}$) and
decrease the estimated mass ($\propto f^{2/3}$). However, adopting
$f=1$ for simplicity, a mass estimate of H\,{\sc i} inside an
homogeneous disk of radius 6 pc is $M_{\rm HI}
\sim2\times10^3$M$_{\odot}$. Such a mass would be enough to supply
material to the source for $<3\times10^7$ years (assuming a radiative
efficiency $\eta<10\%$), given that the total luminosity of the radio
source is $\sim3.6\times10^{41}$ erg/s (e.g.\ Ferrarese et al.\
\cite{ferrarese96}). Using the correlation between FRI source sizes
and their age (Parma et al.\ \cite{parma99}), the size of \4261 (Jaffe
\& McNamara \cite{jaffe94}) implies an age $\sim3\times10^7$
years. The same correlation shows other FRIs with ages $>10^8$ years.
Hence, on this time-scale the H\,{\sc i} mass we estimate is barely
sufficient to fuel the source. It seems more plausible that there is a
continuous flow of accreting material being transported from the 100
pc scale dust disk onto the central nucleus.

The circumnuclear torus- or disk-structures observed in H\,{\sc i} are
usually found on slightly larger scales (50 -- 100 pc; e.g.\ Gallimore
et al.\ \cite{gallimore99} and Conway \cite{conway99}). Only in a few
other cases the H\,{\sc i} is found to lie on very small scales ($<10$
pc in NGC\,4151; Mundell et al.\ \cite{mundell96} and Gallimore et
al.\ \cite{gallimore99}) and it is not obvious that H\,{\sc i}
survives so close to the nucleus. For gas irradiated by X-rays an
effective ionisation parameter $\xi_{\rm eff}$ can be defined, which
governs the physical state of the gas (Maloney et al.\
\cite{maloney96}). For $\xi_{\rm eff}<10^{-3}$ the gas is likely to be
molecular with gas temperatures close to or below 100 K, while higher
values of $\xi_{\rm eff}$ correspond to a hotter atomic gas
phase. Following Maloney et. al\ (\cite{maloney96}), we use $\xi_{\rm
eff}=L_{\rm x}/(r^2nN_{22}^{0.9})$, where $L_{\rm x}$ is the hard
($>1$ keV) X-ray luminosity, $r$ is the distance from the nucleus to
the irradiated gas, $n$ is the gas density and $N_{22}$ is the column
density in units of $10^{22}$ cm$^{-2}$. At the distance of 6 pc a gas
density of $6\times10^2$ cm$^{-3}$ yields an (atomic) obscuring column
density of $N_{22}\sim1.1$. Using the hard X-ray luminosity of \4261
($10^{41}$ erg/s, Roberts et al.\ \cite{roberts91}) this results in
$\xi_{\rm eff}=0.5$; thus implying a mainly atomic gas phase where the
gas temperature is likely to exceed 1000 K (Maloney et al.\
\cite{maloney96}). As a consequence the spin temperature is probably
larger than 100 K, and our estimates of the H\,{\sc i} mass and
density will only be lower limits.

We conclude that within the scope of this model, it is indeed possible
to have an atomic structure on the scales sampled by our VLBI
observations. The inner boundary of this region is naturally set by
the location of the free-free absorption, which also must be
geometrically thin in order to leave the core unattenuated. On the
outside, the structure changes over into a dust disk which is visible
to HST from its innermost pixel, at $r\sim6$ out to 240 pc. However,
since one would think that the mm radiation originates from the flat
spectrum core, it is difficult to reconcile the reported CO absorption
(Jaffe \& McNamara \cite{jaffe94}) with a thin molecular disk. Apart
from the unknown location of the CO gas, the evidence points to the
FRI radio-source in \4261 being powered by gas infall through a
relatively thin disk with a clear gradient of excitation conditions.

\begin{acknowledgements}
  The scientific observations presented in this paper were made
  possible by the dedication and expertise of the teams involved in
  constructing the EVN MkIV data processor and implementing the EVN
  MkIV upgrade at the stations. We acknowledge especially the effort
  required by the correlator team to produce the data for this project
  in a such an early stage.
\end{acknowledgements}

\end{document}